# Measurement of the radiation field surrounding the Collider Detector at Fermilab

Kostas Kordas, Saverio D'Auria, Andy Hocker, Susan McGimpsey, Ludovic Nicolas, Richard J. Tesarek, and Steven Worm
(CDF radiation monitoring group)

***Abstract*— We present here the first direct and detailed measurements of the spatial distribution of the ionizing radiation surrounding a hadron collider experiment. Using data from two different exposures we measure the effect of additional shielding on the radiation field around the Collider Detector at Fermilab (CDF). Employing a simple model we parameterize the ionizing radiation field surrounding the detector.**

## I. Introduction

IN modern collider experiments, the supporting infrastructure lies external to the detector, but inside the radiation environment surrounding the detector. The apparatus and its infrastructure may be sensitive to both chronic and acute radiation doses. These doses induce additional detector occupancy, single-event effects in the supporting electronics, or even irreversible failure. This sensitivity can lead to additional contamination of physics signals, corruption of the data, reduced reliability of the detector, or reduced detector lifetime [1]. Knowledge of the spatial distribution, dose rate and sources of radiation are, therefore, critical components in the design and operation of an experiment at a hadron collider. Most experiment designs have relied on a combination of radiation damage measurements and computer simulations of the radiation environment [2], [3], [4]. However, no substantial measurements of the radiation field surrounding a collider detector exist in the literature.

In this article, we present the first detailed measurement of the radiation field surrounding the Collider Detector at Fermilab (CDF), operating at the Tevatron proton-antiproton collider. We use two types of thermal luminescent dosimeters to measure both the ionizing radiation and the radiation from low energy neutrons and we report here the results for the ionizing radiation field. By comparing the results from two exposure periods, we evaluate the effectiveness of additional shielding installed between the exposures.

Manuscript received October 29, 2003. This work was supported by the U.S. Department of Energy and the National Science Foundation; the UK Particle Physics and Astronomy Research Council; and the Natural Sciences and Engineering Research Council of Canada.

Kostas Kordas is the corresponding author and is with the University of Toronto, Toronto, Ontario M5S 1A7, Canada (e-mail: kordas@fnal.gov)

Saverio D'Auria is with Glasgow University, Glasgow G12 8QQ, United Kingdom (e-mail: dauria@fnal.gov)

Andy Hocker is with the University of Rochester, Rochester, New York 14627, USA (e-mail: hocker@fnal.gov)

Susan McGimpsey and Richard J. Tesarek are with the Fermi National Accelerator Laboratory, Batavia, Illinois 60510, USA (e-mails: mcgimpsey@fnal.gov and tesarek@fnal.gov, respectively)

Ludovic Nicolas is with Glasgow University, Glasgow G12 8QQ, United Kingdom, and with the Fermi National Accelerator Laboratory, Batavia, Illinois 60510, USA (e-mail: nicolas@fnal.gov)

Steven Worm is with Rutgers University, Piscataway, New Jersey 08855, USA (e-mail: worm@fnal.gov)

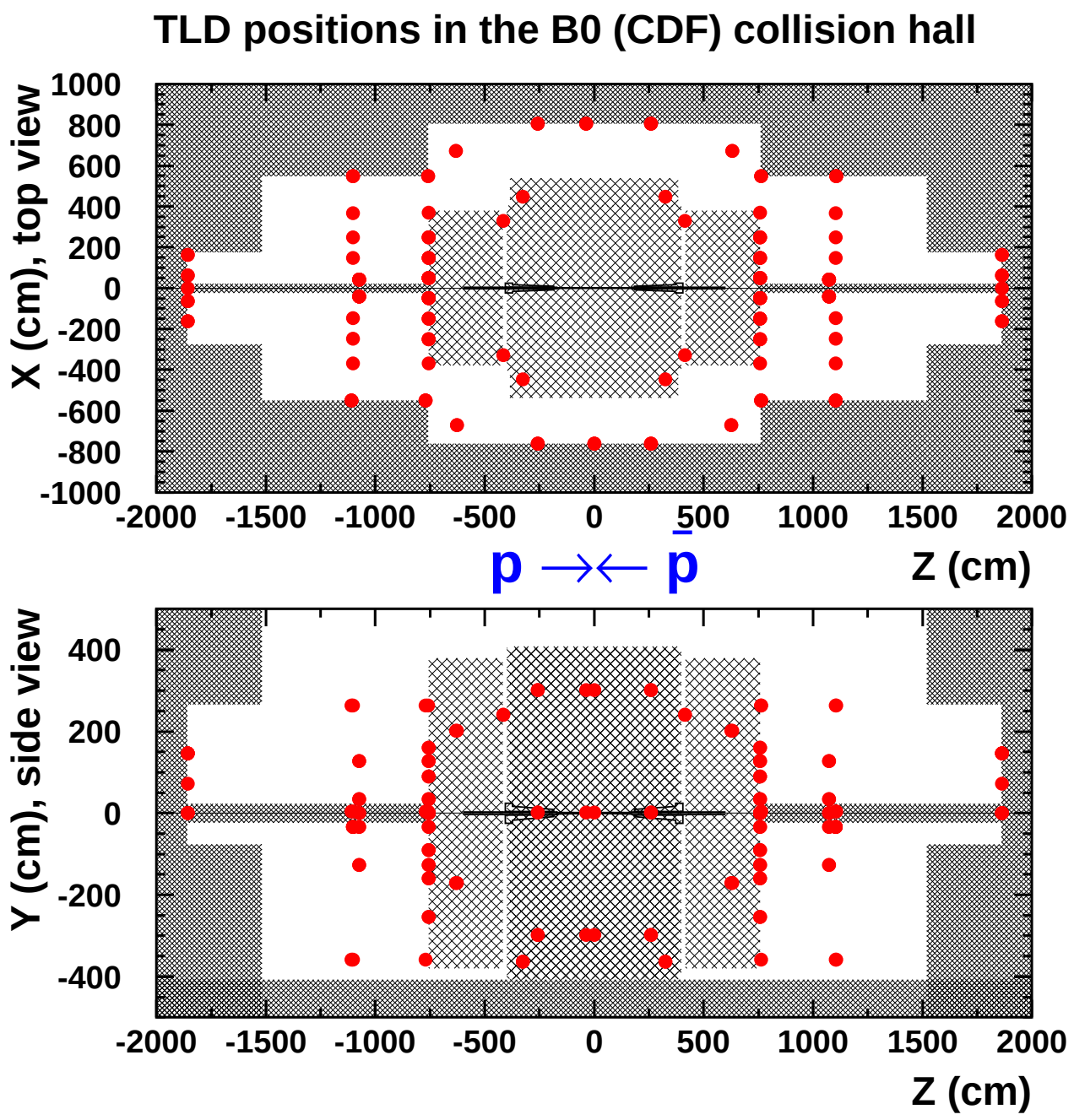

Fig. 1. *Top:* Plane view of the Collider Detector at Fermilab (lightly hatched area) placed at the center of the B0 collision hall (defined by the concrete walls, shown as the heavily hatched area in the figure). Protons ($p$) and antiprotons ($\bar{p}$) can collide at the center of the detector, at $z = 0$. The circles denote the locations where we measure the radiation dose. *Bottom:* Elevation view of the same setup.

## II. Radiation Field Measurement

### A. The Collider Detector at Fermilab

CDF records particles produced in proton-antiproton collisions by means of various detectors surrounding the beam line in a cylindrical geometry. The Tevatron collider provides protons ($p$) and antiprotons ($\bar{p}$) which can collide every ≃396 ns with an energy of 1.96 TeV. Protons travel along the $+z$ direction and collide with oncoming antiprotons at the center of the CDF detector at $z = 0$ (see Fig. 1). In the CDF cylindrical geometry we denote the distance from the beam line by $r$, and the azimuthal angle around the $z$-axis by $\phi$.

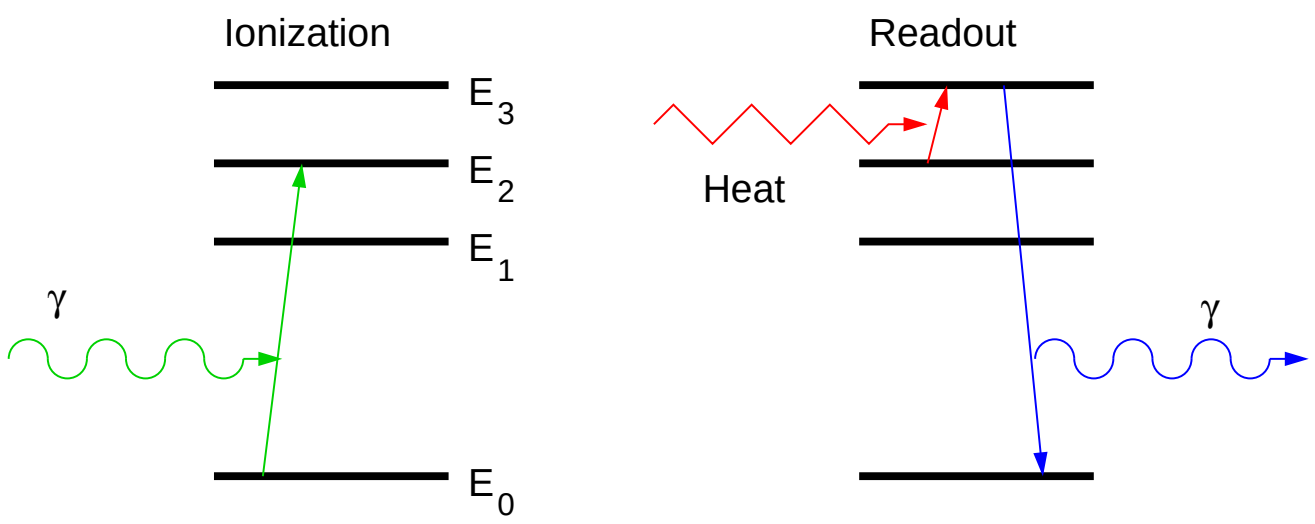

Fig. 2. The principle of thermal luminescence. Photon radiation brings the material in a meta-stable state, $E_2$, with a long lifetime (left). Heating the material leads to emission of visible photons (right).

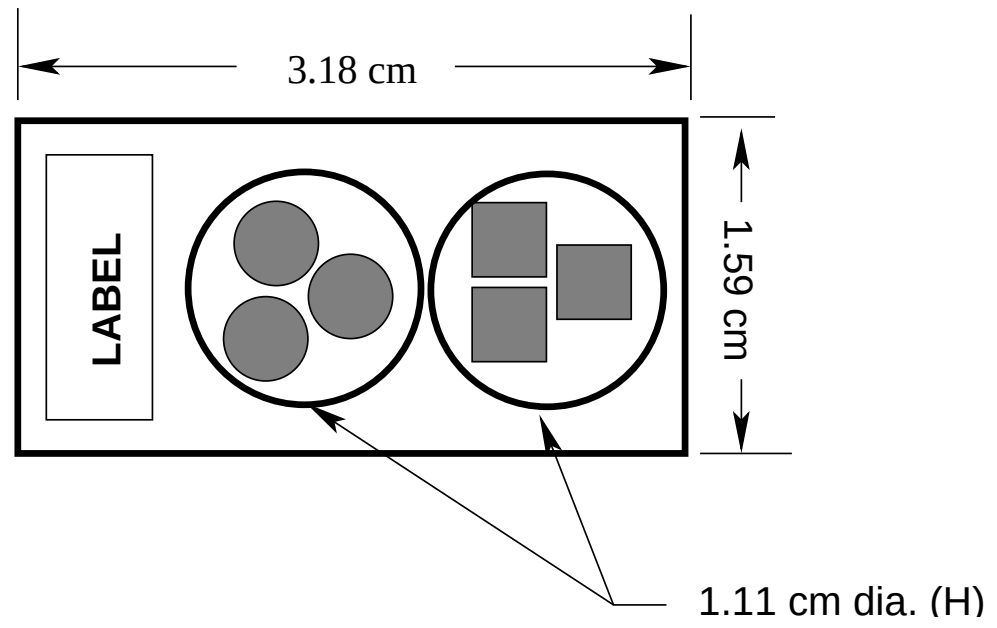

Fig. 3. A 0.79 mm thick FR-4 TLD holder. TLD-700 (round) and TLD-600 (square) dosimeters are kept in place by 76 $\mu$m thick kapton tape.

A series of semiconductor and gaseous detectors, immersed in a 1.4 T solenoidal magnetic field within 1.5 m of the beam line, measure charged particles produced at $p\bar{p}$ collisions. Outside the tracking volume, calorimeters measure the total energy of neutral and charged particles from the proton-antiproton collisions. The calorimeters are surrounded by muon detectors.

The number of $p\bar{p}$ collisions at the center of CDF is recorded by the Cherenkov Luminosity Counter (CLC) [5]. On either side of the detector, scintillator counters surrounding the beam pipe record losses from protons and antiprotons ejected from the beam. Proton (antiproton) losses are defined as the coincidence of a counter signal with a proton (antiproton) bunch crossing the plane of the scintillator on its way into the CDF detector.

### B. Thermal Luminescent Dosimeters

Two types of Harshaw TLD chips are used for the radiation measurements. One type (TLD-700) is based on $^7$LiF and is sensitive to ionizing radiation. Ionizing radiation passing through the dosimeter brings the material in a meta-stable state with very long lifetime. Heating the TLD chip leads to a transition back to ground state accompanied by the emission of a photon (see Fig. 2). The number of photons produced is proportional to the population in these meta-stable states, which is in turn proportional to the amount of ionizing radiation that has traversed the TLD chip. The other dosimeter type (TLD-600) is based on $^6$LiF and is sensitive to both ionizing radiation and low-energy neutrons ($E_n < 200$ keV). The reaction $n + {}^6\mathrm{Li} \rightarrow {}^3\mathrm{H} + \alpha$ results in a transition to the meta-stable state discussed above, by means of the recoiling tritium ($^3$H) and helium ($\alpha$) nuclei.

Dosimeters are grouped in two triplets, one of each TLD type, and put in 3.18 cm $\times$ 1.59 cm holders made of 0.79 mm thick FR-4 (see Fig. 3). The TLD's are held in place by 76 $\mu$m thick kapton tape, and are subsequently placed in 160 locations around the collision hall to accumulate radiation, on both the proton ($z < 0$) and the antiproton ($z > 0$) sides (see Fig. 1): i) around the entrance points of the beams to the collision hall, at 9 locations on each side, at $z = \pm 1860$ cm, $60 < r < 220$ cm, $0 < \phi < \pi$, ii) on the horizontal and vertical bars supporting the Tevatron quadrupoles, at 16 locations on each side, at $z \simeq \pm 1090$ cm, $40 < r < 660$ cm, $0 < \phi < 2\pi$, iii) on the face of the steel wheels hosting the forward muon detectors, at 26 locations on each side, at $z = \pm 757$ cm, $49 < r < 610$ cm, $0 < \phi < 2\pi$, iv) on the collision hall walls running parallel to the beam line, at 42 locations, at $z = -1105, -760, -259, 0, 259, 760, 1105$ cm, $540 < r < 860$ cm, $0 < \phi < 2\pi$, v) on the racks hosting readout electronics for the silicon tracking detectors, at 8 locations, at $z = -630, 630$ cm, $r \simeq 695$ cm, $0.3 < \phi < 6.0$, and vi) on the racks hosting power supplies for the drift chamber tracker, at 8 locations, at $z = -416, -326, 326, 416$ cm, $325 < r < 416$ cm, $0.6 < \phi < 5.6$.

### C. Calibration and Dosimetry

We calibrate the TLD response to ionizing radiation with a 1 rad photon exposure from a $^{137}$Cs source [6]. A calibration factor (in rad/nC) for each TLD chip is then determined by heating up the chip and measuring the light yield using a Harshaw model 2000 TLD reader [7]. A reproducibility of $\sim$1% and a chip-to-chip variation of $\sim$3% is observed. The response of the TLD-600 chips to neutrons is calibrated with a 1 rad exposure to a $^{252}$Cf source. We obtain a $\sim$10% reproducability and a $\sim$15% chip-to-chip variation.

LiF TLD's are known to exhibit non-linearity for doses above 100 rad. In order to account for this behavior, we expose a small sample of TLD's to doses up to 10 krad and we measure a correction factor, defined as the ratio of the received dose over the dose estimated from the linear-response assumption (see Fig. 4). The dosimeters exposed around the CDF collision hall have measured doses in the range of 0.1 rad to 1.2 krad.

We extract the ionizing radiation, $D_\gamma$ (rad), each TLD-700 chip has received due to its exposure in the collision hall, by using the expression:

$$D_\gamma = C_{\gamma,7}\, k_{\gamma,7}\, R_7 - D_{\gamma,ctrl} \;, \qquad (1)$$

where $R_7$ is the reading (nC) from this TLD chip, $k_{\gamma,7}$ is the calibration factor (rad/nC) for its response to ionizing radiation,

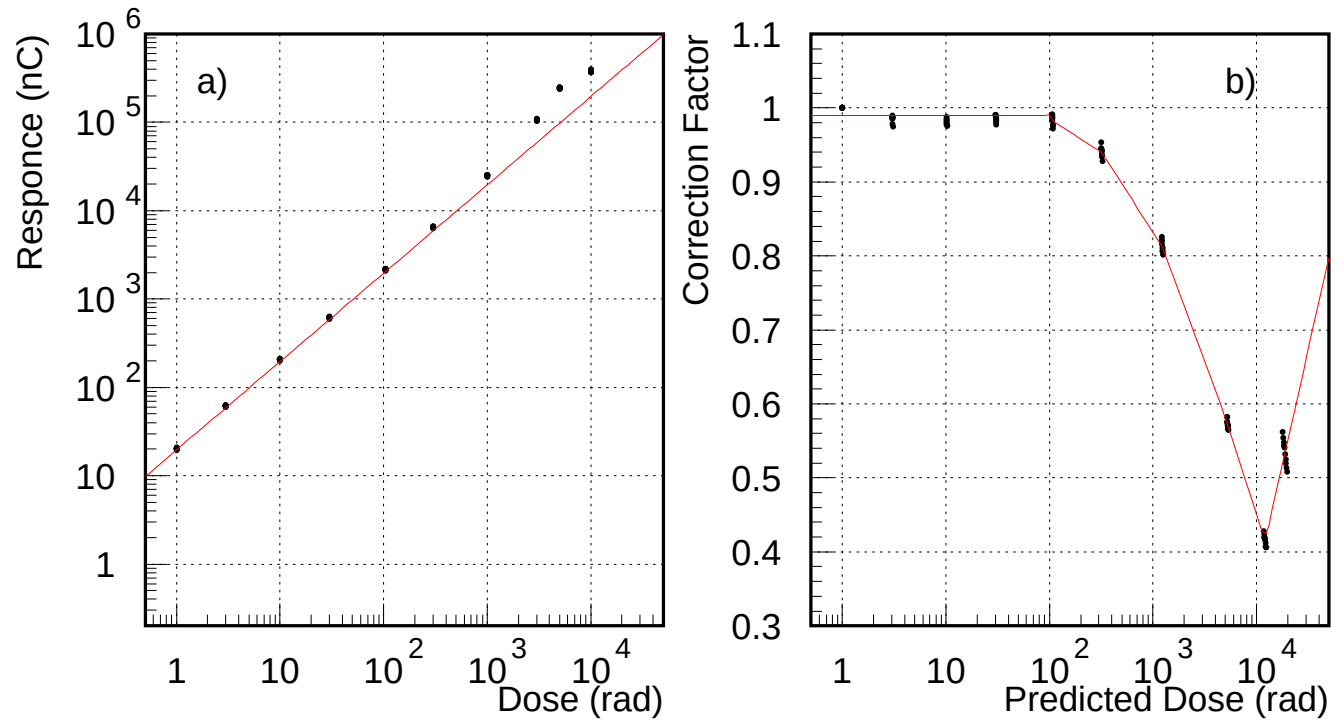

Fig. 4. a) Response of TLD-700 dosimeters to ionizing radiation as a function of received dose; note the super-linear behavior for doses above 100 rad. b) The non-linearity correction factor as a function of the dose estimated from the linear-response assumption.

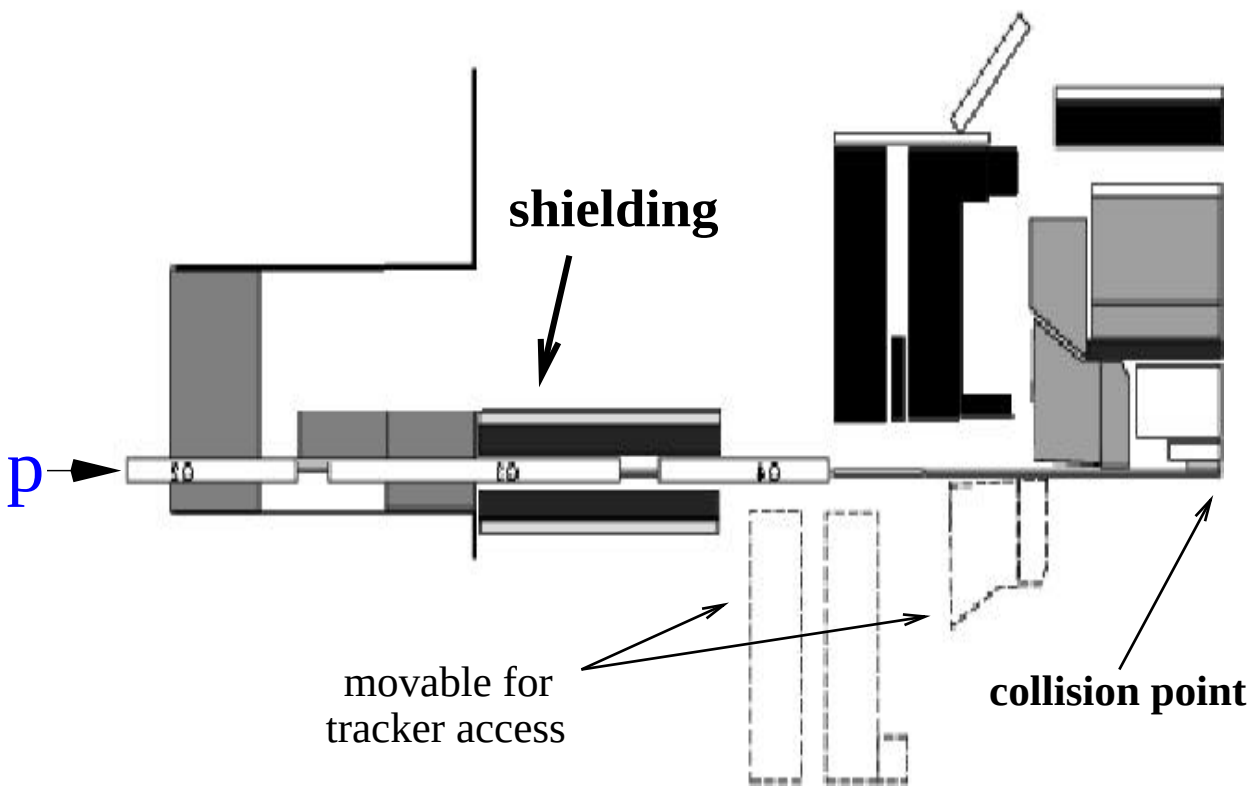

Fig. 5. Elevation view of a quadrant of CDF, with the shielding installed around the focusing quadrupoles on the proton side, just before data-taking resumed at the end of January 2003 (beginning of exposure period 3).

TABLE I

BEAM CONDITIONS AT CDF FOR THE THREE TLD EXPOSURE PERIODS.

| Period | Beam ($\times 10^{18}$) $p$ | $\bar{p}$ | Losses ($\times 10^{9}$) $p$ | $\bar{p}$ | $\int Ldt$ (pb$^{-1}$) |
|---|---|---|---|---|---|
| 1) May - Jun. 2002 | 4.34 | 0.19 | 8.16 | 1.41 | 5.49 |
| 2) Jun. - Oct. 2002 | 31.7 | 1.92 | 80.1 | 11.3 | 56.4 |
| 3) Jan. - May 2003 | 29.4 | 2.32 | 61.5 | 7.5 | 74.8 |

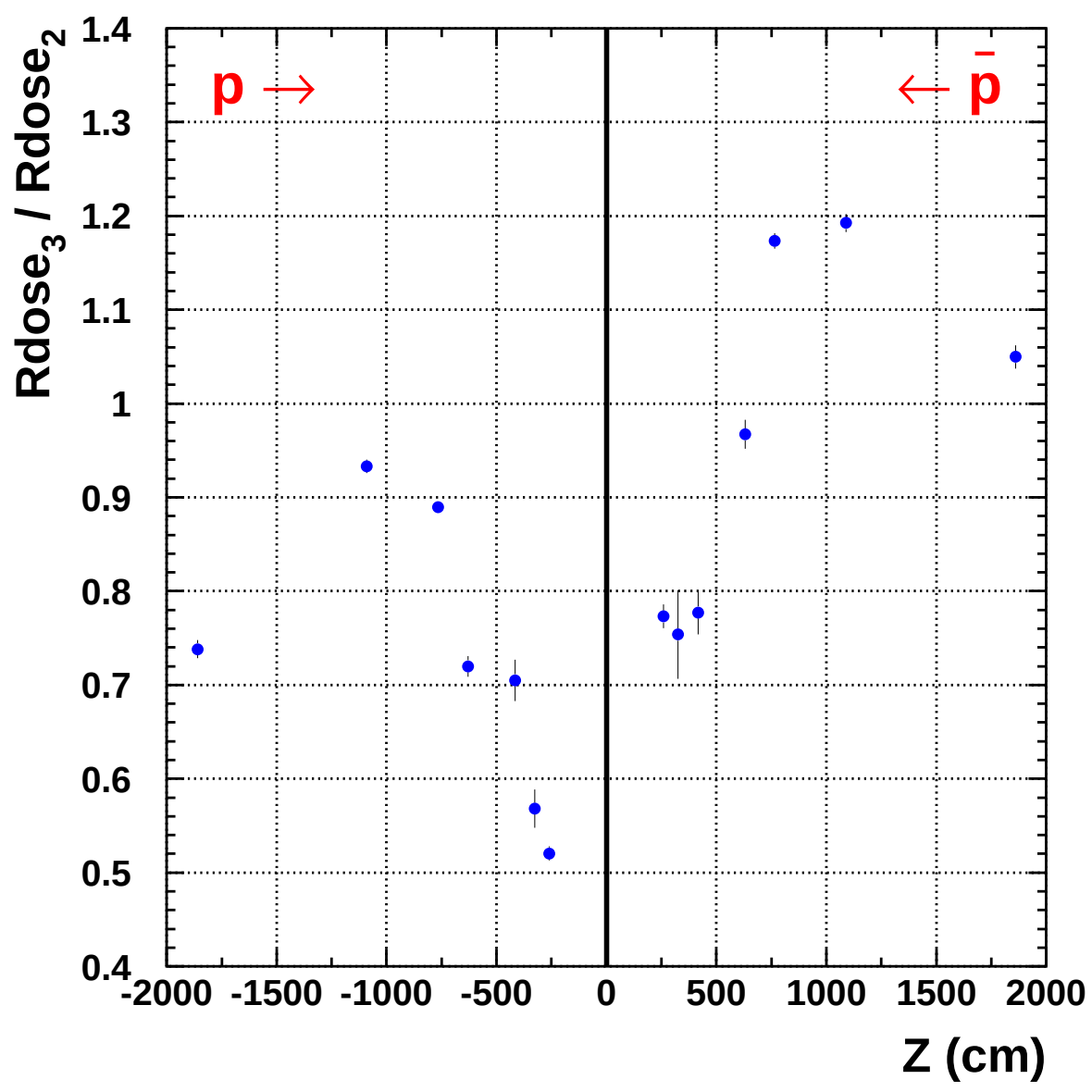

Fig. 6. Ratio (period 3 over period 2) of ionizing radiation dose rates, $Rdose$ (dose per $pb^{-1}$ of collisions), at various $z$ locations.

$C_{\gamma,7}$ is the non-linearity correction factor, and $D_{\gamma,ctrl}$ is the background ionizing radiation dose measured by a number of control TLD-700 chips which were not placed in the collision hall. Averaging the doses measured by the three TLD chips in a given holder, we obtain the ionizing radiation dose, $D_\gamma$, at the location of the TLD holder in study.

### *D. Radiation measurements and effectiveness of shielding*

TLD measurements are taken during three different periods of the Tevatron operations. Table I shows the integrated beam conditions during the three exposure periods: the number of protons and antiprotons in the Tevatron, the number of lost beam particles recorded, and the number of collisions in terms of time-integrated luminosity, $\int Ldt$ ($1\ pb^{-1}$ corresponds to about $5 \times 10^{10}$ $p\bar{p}$ interactions). The first exposure period was a test period; only a partial set of TLDs was installed around the collision hall. We, therefore, focus our discussion to period 2 (June to October 2002) and period 3 (January to May 2003).

During a break in the Tevatron operations in January 2003 (just before period 3 commenced), shielding was installed around the focusing quadrupoles on the proton side (see Fig. 5). No shielding was installed on the antiproton side because the beam losses are much smaller (see Table I).

In Figure 6 we show the ratio of the dose rate, $Rdose$ (dose per $pb^{-1}$ of collisions), in period 3 over that of period 2 at various $z$ locations. Each point on the plot is the weighted average of the ratios of the measurements in $r$ and $\phi$ for the given $z$ location. On the antiproton side, where no shielding was installed, the dose rates in period 3 are not consistently higher or lower compared to period 2; the dose rates range from $\sim$20% higher (at $z \simeq 1090$ cm), to $\sim$22% lower (at $z \sim 300$ cm). On the proton side, where shielding was installed, the dose rates in period 3 are systematically lower than in period 2; from $\sim$6% (at $z \simeq -1090$ cm) to $\sim$48% (at $z = 259$ cm), for an average reduction of $\sim$25%.

Assuming that the radiation at a given point is the linear super-position of contributions from beam losses and collisions, we can write the dose rate, $Rdose$, as

$$Rdose \equiv \frac{D}{\int Ldt} = \frac{D_C}{\int Ldt} + \frac{D_L}{\int Ldt}, \tag{2}$$

where $D$, $D_C$ and $D_L$ denote the measured dose, the dose due to collisions and the dose due to beam losses, respectively.

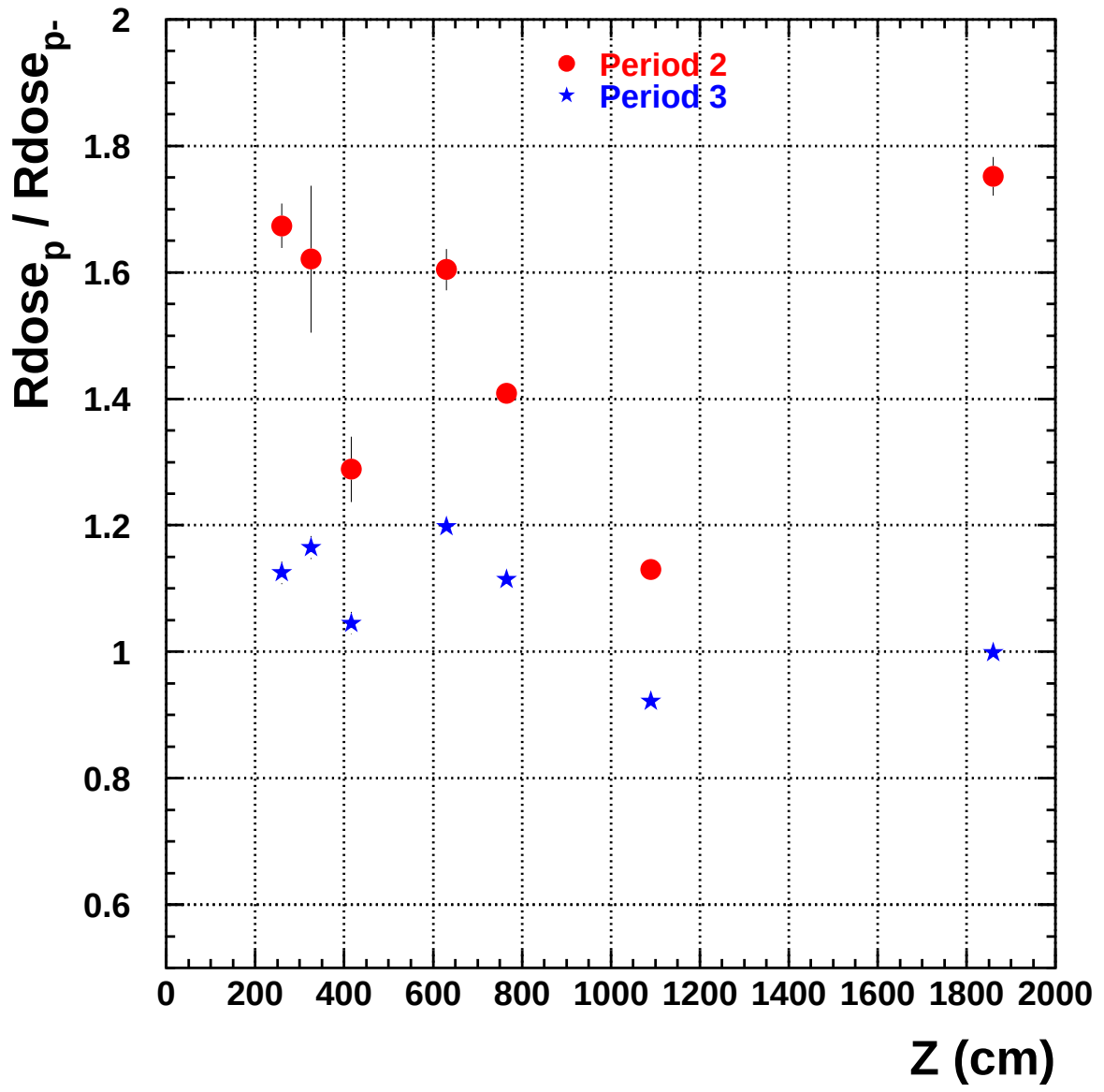

Fig. 7. Ratio of dose rates on the proton side over the antiproton side, at various $z$ locations in exposure periods 2 (circles) and 3 (stars).

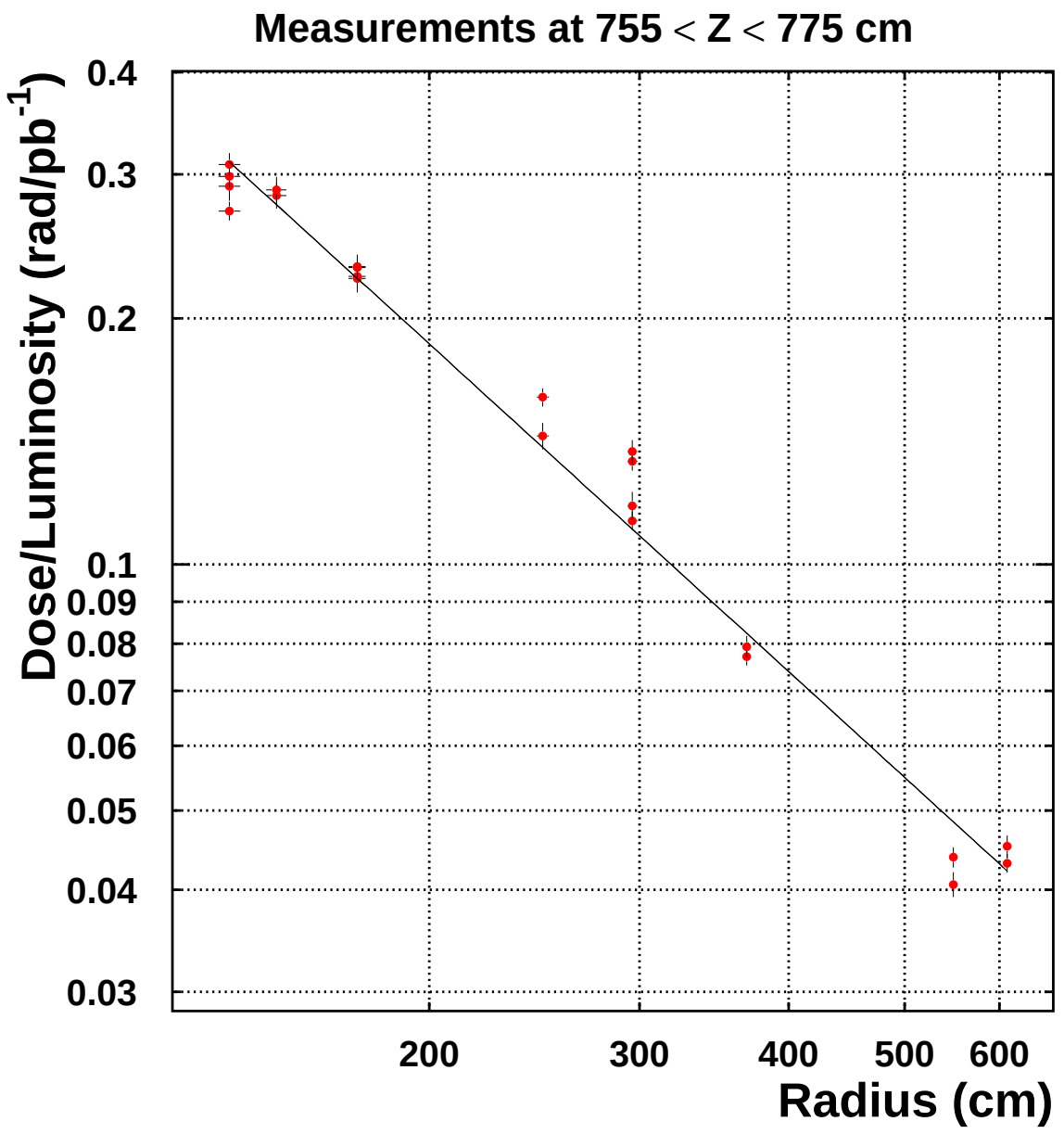

Fig. 8. Dose rate (dose per $pb^{-1}$ of collisions) as a function of the distance from the beam line, for measurements at $z \simeq 760$ cm in exposure period 2. The data are fitted to the radiation filed model in Equation 3.

If we assume that the collision contribution to the dose ($D_C$) scales with the number of collisions, we expect that $D_C / \int Ldt$ is the same in periods 2 and 3 at the points where we perform our measurements. The fact that the dose rates are different in period 3 than in period 2 means that the rate of the loss contributions ($D_L / \int Ldt$) is different in the two periods (see Eqn. 2). Therefore, we conclude that the $\sim$25% reduction in dose rates on the proton side, quoted in the previous paragraph, is solely due to a reduction in the beam loss rates.

Figure 7 shows the ratio of the dose rates on the proton and antiproton sides, at several $z$ locations in periods 2 (circles) and 3 (stars). In both exposure periods, the dose rates on the proton side are usually higher than those on the antiproton side. In period 2 asymmetries as high as 80% are observed, whereas in period 3, when the shielding on the proton side was installed, this asymmetry is no more than 20%. Given the symmetry of the CDF detector, we can assume that the dose contribution due to collisions does not exhibit a preference for positive $z$ values over negative $z$ values. Thus, we expect the dose rate asymmetry between the proton and antiproton sides to arise from an asymmetry in the rate of loss contributions (see Eqn. 2).

## III. Modeling the Radiation Field

The ionizing radiation measurements are parameterized using a model based on previous CDF measurements of the silicon radiation damage profile [8] and direct radiation measurements in the CDF tracking volume [9]. This model assumes cylindrical symmetry of the radiation around the beam line, with a radial dependence which follows a power law in $1/r$, where $r$ is the distance from the beam line. For any point $(x, y)$ on a plane perpendicular to the beam axis at $z$, we write for the dose rate (dose per $pb^{-1}$ of collisions), $Rdose(r)$:

$$Rdose(r) = Ar^{-\alpha}, \quad (3)$$

where $A$ is the absolute normalization, $\alpha$ is the power law exponent, and $r = \sqrt{x^2 + y^2}$ is the distance from the beam line ($z$-axis). An example is shown in Fig. 8 for measurements on the antiproton side, at $z \simeq 760$ cm in exposure period 2.

The resulting normalizations and power law exponents are shown in Figures 9 and 10 for measurements in periods 2 and 3, respectively. For the region $760 < |z| < 1090$ cm, outside the CDF main volume, the radiation field behaves similarly between the proton and antiproton sides; the normalizations are of the same order of magnitude and the power law exponents have values $\alpha \sim 1.3 - 1.4$ The situation is quite different around the entrance points of the beams to the collision hall, at $z = \pm 1860$ cm. The radiation on the proton side has a much higher normalization than on the antiproton side, but it decreases much faster with the distance from the beam.

The power law of Equation 3, with amplitudes and exponents from Fig. 9 or Fig. 10, can be used to predict the radiation level at any point surrounding the CDF detector.

## IV. Conclusion

Using Thermal Luminescent Dosimeters (TLD's) we have presented a detailed measurement of the ionizing radiation field around the Collider Detector at Fermilab (CDF). Using data from two exposures we are able to evaluate the effectiveness of shielding installed on the proton side to reduce the radiation levels. When a simple model assuming a power law in $1/r$ is used, fits to the data yield exponents with a strong $z$-dependence; $0.7 - 1.8$ in the collision hall hosting the CDF detector.

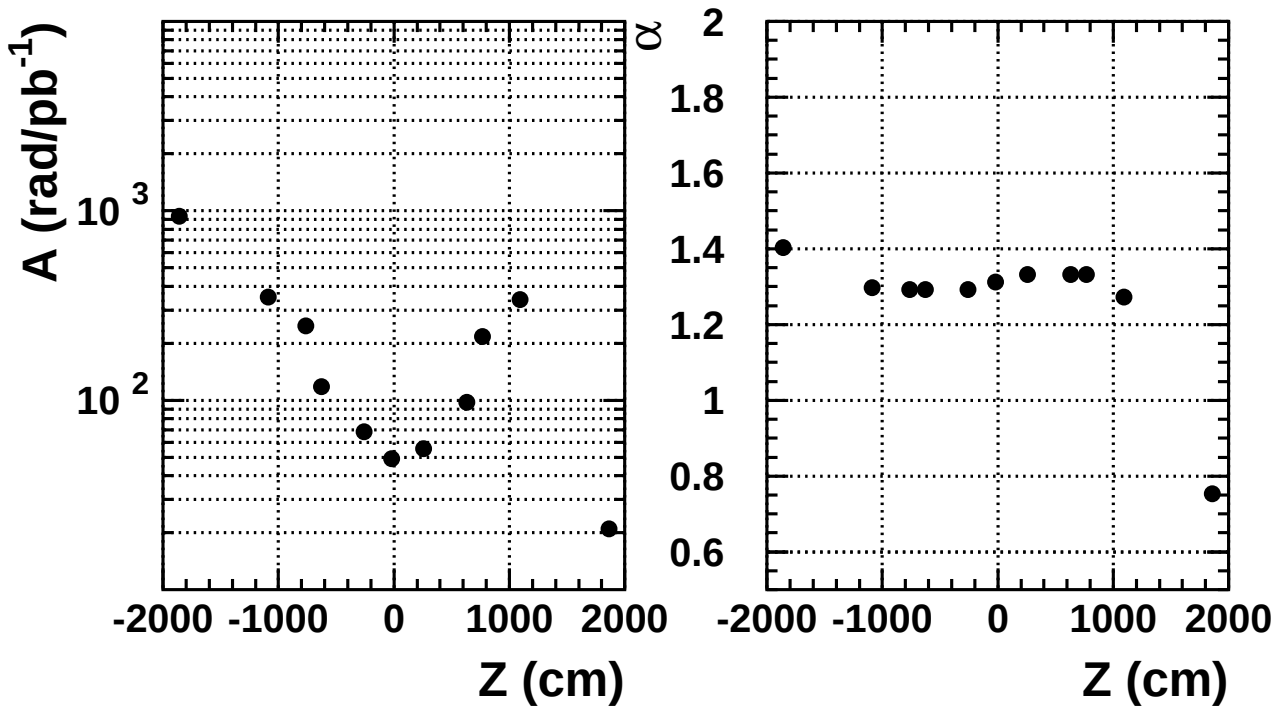

Fig. 9. Fit parameters of the radiation field model in Equation 3 ($Rdose(r) = A\ r^{-\alpha}$), for measurements in period 2; normalization (left) and power law exponent (right) as a function of $z$.

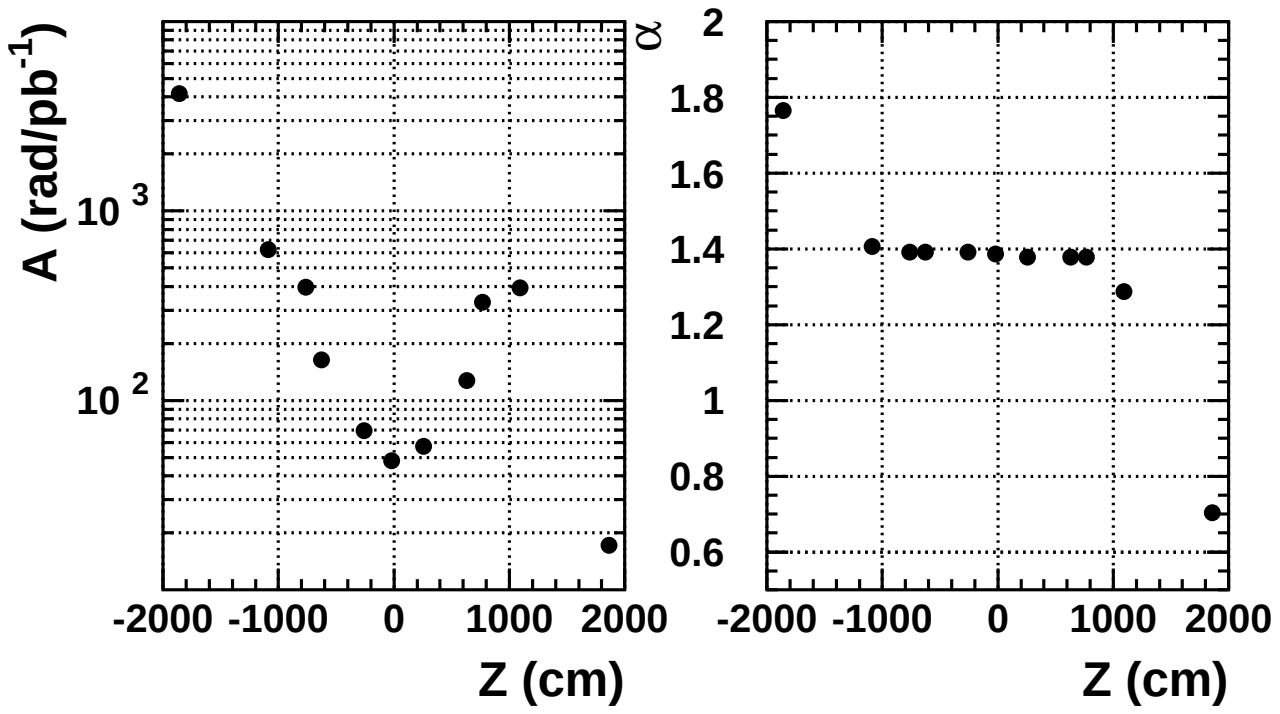

Fig. 10. Fit parameters of the radiation field model in Equation 3 ( $Rdose(r) = A\ r^{-\alpha}$ ), for measurements in period 3; normalization (left) and power law exponent (right) as a function of $z$.

We believe that our data can serve as a calibration point for simulations of the radiation environment in future hadron colliders.

## Acknowledgment

The authors would like to thank the people at Fermilab's Radiation Physics Calibration Facility, the Silicon Detector Lab and CDF for their help in calibrating the TLD's, placing them to the holders and installing them in CDF. Special thanks to Minjeong Kim and Fabio Happacher for the invaluable help in placing/harvesting the dosimeters, and to the radiation monitoring group at Argonne labs for letting us use their TLD reader when needed.